\documentclass[12pt]{article}
\usepackage{amsfonts,epsfig,latexsym,amscd,amsmath,theorem,mathrsfs,graphics}

\DeclareMathOperator{\diag}{diag}

\newcommand{\Rvec}{{\bf R}} 
\newcommand{\Svec}{{\bf S}}

\newcommand{\xv}{{\bf x}}

\newcommand{\R}{{\mathbb{R}}}
\newcommand{\Z}{{\mathbb{Z}}}

\newcommand{\C}{{\mathbb{C}}}
\newcommand{\I}{{\mathbb{I}}}
\newcommand{\LL}{{\mathscr{L}}}
\newcommand{\MM}{{\mathscr{M}}}

\newcommand{\beq}{\begin{equation}}
\newcommand{\eeq}{\end{equation}}
\newcommand{\bea}{\begin{eqnarray}}
\newcommand{\eea}{\end{eqnarray}}
\newcommand{\ben}{\begin{eqnarray*}}
\newcommand{\een}{\end{eqnarray*}}
\newcommand{\bem}{\begin{enumerate}}
\newcommand{\eem}{\end{enumerate}}

\newcommand{\ra}{\rightarrow}

\newcommand{\cd}{\partial}
\newcommand{\wt}{\widetilde}

\newcommand{\su}{{\mathfrak{su}}}

\def \d{\mathrm{d}}

\newcommand{\ignore}[1]{}

\newcommand{\RR}{\mathscr{R}}

\newcommand{\pivec}{\mbox{\boldmath{$\pi$}}}

\newcommand{\tauvec}{\mbox{\boldmath{$\tau$}}}

\newcommand{\nvec}{\mbox{\boldmath{$n$}}}

\newcommand{\Xvec}{\mbox{\boldmath{$X$}}}

\newcommand{\tr}{{\rm tr}\, }

\theoremstyle{plain}

{\theorembodyfont{\rmfamily}

}

\newcommand{\news}{\setcounter{equation}{0}}

\begin{document}

\title{A simple mass-splitting mechanism in the Skyrme model}
\author{J.M. Speight\thanks{E-mail: {\tt speight@maths.leeds.ac.uk}}\\
School of Mathematics, University of Leeds\\
Leeds LS2 9JT, England}

\maketitle

\begin{abstract}
It is shown that the addition of a single chiral symmetry breaking term to the standard
$\omega$-meson variant of the nuclear Skyrme model can reproduce the proton-neutron mass difference.
\end{abstract}

\maketitle

\section{Introduction}
\news

The Skyrme model \cite{sky} is a nonlinear sigma model whose perturbative quanta are interpreted as pions and whose
topological solitons are interpreted as nucleons. At the classical level, there is no difference between
protons and neutrons in this model; the distinction only emerges after quantization of the internal rotational
degrees of freedom enjoyed by the static soliton, and is determined, roughly speaking, by the sense (clockwise or anticlockwise) of internal rotation. It is clear, therefore, that in order to account for the slight mass difference between the neutron and the proton ($m_n= 939.56563$ MeV, $m_p= 938.27231$ MeV), the model's action must contain a small term which breaks time reversal symmetry, so that ``clockwise" and ``anticlockwise" internal rotation are dynamically inequivalent. Devising such a term, while maintaining Lorentz and parity invariance, does not seem to be possible in any version of the model containing only pion fields. It should be noted that adding electromagnetic effects to the
usual Skyrme model has precisely the opposite effect to the one desired: unsurprisingly, Coulomb repulsion renders the proton {\em heavier} than the neutron \cite{ebrsav}. 

As has been long known, the proton-neutron mass splitting can be accommodated if one extends the model to include coupled {\em vector} mesons \cite{jaijohparsch}. There is independent motivation for this: coupling the Skyrme field to the $\omega$ meson stabilizes the unit soliton without the need for a quartic term in the action \cite{adknap}, producing a mathematically elegant, but comparatively underexplored, model whose static solitons are thought to be
qualitatively similar to those of the standard Skyrme model \cite{sut-vecmes}. Another motivation comes from holography. Variants of the Skyrme model coupled to an infinite tower of vector mesons emerge from Yang-Mills theory in
$4+1$ dimensions in both the Sakai-Sugimoto model \cite{saksug} and Sutcliffe's simplification of it \cite{sut-holographic}, which, it has been recently argued, can correct the binding energy problem suffered by the conventional Skyrme model \cite{naysut}. By introducing an explicit mass difference between up and down quarks into the Sakai-Sugimoto model, Bigazzi and Niro have recently
computed the proton-neutron mass splitting contributed by the whole infinite tower of vector mesons, obtaining
a phenomenologically satisfactory answer \cite{bignir}. This result lies at the extreme end of the spectrum of sophistication, requiring, as it does, dynamical coupling to infinitely many new meson fields (though, of course, holography handles them elegantly). Earlier proposals, motivated directly by chiral perturbation theory, require the addition of $\omega$, $\rho^0$ and $\rho^\pm$ mesons \cite{epefangarmen} or $\omega$, $\rho^0$, $\rho^\pm$ and
$\eta$ mesons \cite{jaijohparsch}, producing models of great complexity, in which the full consequences of the various couplings are difficult to apprehend.

The purpose of this paper is to point out that a very simple perturbation of the usual $\omega$ meson version of the Skyrme model \cite{adknap}, containing only one symmetry breaking term and no extra mesons (beyond the $\omega$), can reproduce the proton-neutron mass splitting. A full non-perturbative computation would require one to solve the static Euler-Lagrange equations in the presence of axial, but not spherical, symmetry, a computationally intensive problem. However, working perturbatively in the small coefficient in front of the symmetry breaking term, the leading order
correction can be computed by solving only ODEs. The skyrmion that emerges is (to this order) still a spherically symmetric hedgehog with a radial $\omega_0$ component, but acquires a small azimuthal spatial $\omega$ component localized around the skyrmion's equator. The action induced by rigidly isorotating this skyrmion depends on the sense in which it spins, due to both the small $\omega_i$ field and direct coupling between the $\omega_0$ and pion fields.

\section{The model and its skyrmion}
\news

The usual $\omega$ meson variant of the Skyrme model \cite{adknap} has Lagrangian density, in appropriate
length and energy units \cite{sut-vecmes},
\beq
\LL=\frac{1}{16}\tr(\cd_\mu U\cd^\mu U^\dagger)+\frac{M^2}{8}\tr(U-\I_2)-\frac14\omega_{\mu\nu}\omega^{\mu\nu}
+\frac12\omega_\mu\omega^\mu+\beta\omega_\mu B^\mu
\eeq
where $U:\R^{3,1}\ra SU(2)$ is the Skyrme field, $R_\mu=\cd_\mu U\, U^\dagger$ is its right-invariant
current,
\beq
B_\mu=\frac{1}{3!}\frac{1}{2\pi^2}\epsilon_{\mu\nu\alpha\beta}\frac12\tr(R^\nu R^\alpha R^\beta),
\eeq
is its baryon current,
$\omega=\omega_\mu dx^\mu$ is a real one-form, representing the 
$\omega$ meson, $\omega_{\mu\nu}=
\cd_\mu\omega_\nu-\cd_\nu\omega_\mu$, $M=m_\pi/m_\omega=0.176$ is the ratio of the pion mass to the $\omega$ mass, and $\beta$ is an unknown coupling constant. Adkins and Nappi propose \cite{adknap} the value $\beta=96.7$, found by fitting the masses of the nucleon and $\Delta$ resonance, while Sutcliffe  suggests \cite{sut-vecmes} the alternative $\beta=34.7$, found by fitting the masses of the nucleon and $\alpha$ particle. We prefer the latter procedure, because fitting to a broad resonance like the $\Delta$ seems inherently 
over-optimistic, so will present numerics in this case only.

To $\LL$ we will add the symmetry breaking term
\beq
\LL_*=-\frac\kappa4\omega^{\mu\nu}\Pi_{\mu\nu}
\eeq
where $\kappa$ is a small parameter,
$\Pi_{\mu\nu}=\cd_\mu\pi_1\cd_\nu\pi_2-\cd_\nu\pi_1\cd_\mu\pi_2$ and, as usual, $U=\sigma\I_2+i\tau_a\pi_a$,
$\tau_a$ being the Pauli matrices. This breaks the symmetry of $\LL$ under the adjoint action of $SU(2)$,
$U\mapsto AUA^\dagger$, to invariance under the $U(1)$ subgroup $A=\exp(i\alpha\tau_3)$, but does not alter the pion or 
$\omega$ masses, and is invariant under the parity operation $(t,\xv)\mapsto (t,-\xv)$, 
$U\mapsto U^\dagger$. The Euler-Lagrange equations satisfied by a static solution are
\bea
\frac14\cd_i \Rvec_i-\frac{M^2}{4}\pivec+\frac\kappa2\left\{(\sigma,-\pi_3,\pi_2)\cd_i\pi_2-(\pi_3,\sigma,-\pi_1)
\cd_i\pi_1\right\}\cd_j\omega_{ji}&=&0\nonumber\\
-\cd_j\cd_j\omega_0+\omega_0+gB_0&=&0\nonumber\\
\cd_j\cd_j\omega_i-\cd_i(\cd_j\omega_j)-\omega_i+\frac\kappa2\cd_j\Pi_{ji}&=&0\label{pde}
\eea
where $R_i=i\Rvec_i\cdot\tauvec$. From this we see that $\omega_i$ is of order $\kappa$, and hence $U$ and $\omega_0$ receive corrections only at order $\kappa^2$. Hence, to order $\kappa$, the $B=1$ static solution takes the form
\beq
U=\cos f(r)\I_2+i\sin f(r)\nvec\cdot\tauvec,\quad
\omega_0=\omega_0(r),\quad
\omega_i dx^i=\kappa\bar\omega
\eeq
where
\bea
f''+\frac2rf'-\frac1{r^2}\sin 2f-M^2\sin f+\frac{4g}{2\pi^2r^2}\sin^2f\omega_0'(r)&=&0,\label{o1}\\
\omega_0''+\frac2r\omega_0'-\omega_0+\frac{g}{2\pi^2r^2}f'\sin^2f&=&0,\label{o2}\\
*\d*\d\bar\omega+\bar\omega&=&-\frac12*\d*(\d\pi_1\wedge\d\pi_2),\label{jengre}
\eea
$\pi_1=\sin f(r)\sin\theta\cos\phi$, $\pi_2=\sin f(r)\sin\theta\sin\phi$, and $*$ denotes the Hodge isomorphism
on $\R^3$. Equation (\ref{jengre}) supports solutions of the form
\beq
\bar\omega=\Omega(r)\sin^2\theta \, \d\phi,
\eeq
where
\beq
\Omega''-\left(1+\frac2{r^2}\right)\Omega=\frac18\left(F''-\frac1{r^2}F+\frac1{r^2}\right), \quad
F:=\cos 2f.\label{o3}
\eeq
So constructing the $B=1$ skyrmion to order $\kappa$ amounts to solving the coupled ODE system (\ref{o1}),
(\ref{o2}), (\ref{o3}) with boundary conditions $f(0)=\pi$, $\omega_0'(0)=\Omega(0)=f(\infty)=\omega_0(\infty)=\Omega(\infty)=0$.

\begin{figure}[htb]
\begin{center}
\begin{tabular}{cc} (a) & (b) \\
\includegraphics[scale=0.32]{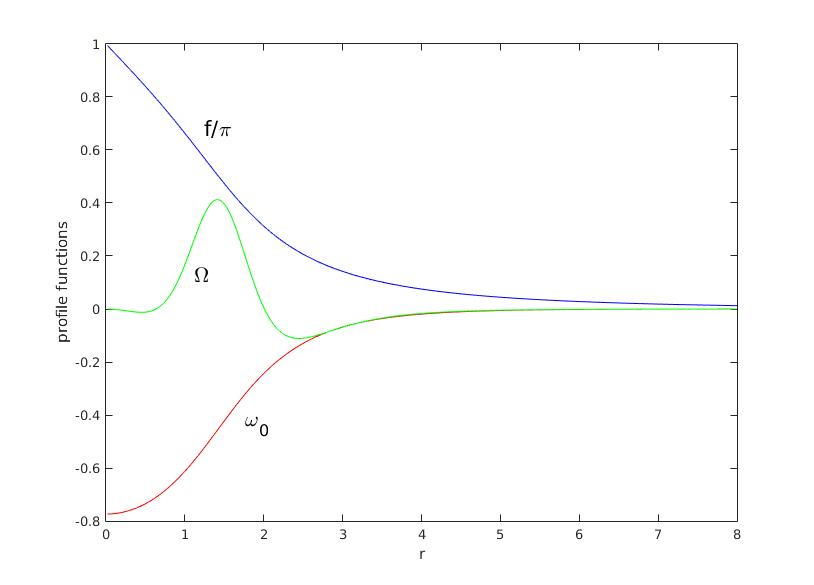}&
\includegraphics[scale=0.45]{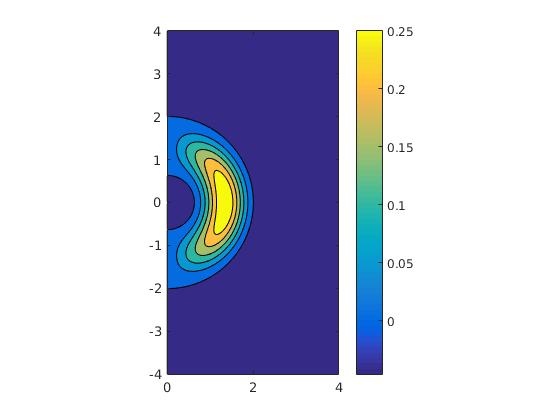}
\end{tabular}
\end{center}
\caption{The perturbed skyrmion: (a) profile functions, and (b) a contour plot of the azimuthal
spatial field $\bar{\omega}$ in the right $x$-$z$ half plane (contours of constant $\bar{\omega}\cdot(0,1,0)$).}
\label{fig1}
\end{figure}

Figure \ref{fig1} presents numerical solutions of this system generated using a relaxation method, for the
coupling value $\beta=34.7$. Note that the Skyrme and $\omega_0$ fields are unchanged by the perturbation
to leading order. Its only effect is to induce a small azimuthal spatial $\omega$ field,
$\omega_i dx^i=\kappa \Omega(r)\sin^2\theta \d\phi$, also depicted.  This perturbed hedgehog solution is, of course,
not unique, since we may act on it by spatial translations and rotations, and isorotations about the
$\pi_3$ axis.

\section{Rigid body quantization}
\news

As usual, we quantize the unit skyrmion by restricting the model's action functional to the symmetry
orbit
of a fixed static skyrmion (neglecting translations, that is, having fixed the skyrmion's centre of mass at the origin). Note that only
isorotations about the $\pi_3$ axis are symmetries of the model.
The spin-isospin symmetry group $G=SU(2)\times U(1)$ acts (on the right) on the space of field configurations by
\beq
(g,\lambda):(U,\omega)\mapsto(U,\omega)_{(g,\lambda)}:=(h(\lambda)^\dagger U(\RR_g\xv)h(\lambda),
\omega_0 dt+\RR_g^*(\omega_i dx^i)),
\eeq
where $h(\lambda):=\diag(\lambda,\bar\lambda)$, and $\RR_g:\R^3\ra\R^3$ denotes the orthogonal linear map corresponding to $\su(2)\ra\su(2)$, $X\mapsto gXg^\dagger$, under the standard identification $\R^3\equiv\su(2)$ defined by $\Xvec\mapsto\Xvec\cdot\left(-\frac{i}{2}\tauvec\right)$.
The $G$ orbit $\MM$ of our reference skyrmion
\beq
(U_*,\omega_*)=(\cos f(r)\I_2+i\sin f(t)\nvec\cdot\tauvec,\omega_0(r)dt+\kappa\Omega(r)\sin^2\theta\d\phi)
\eeq
is diffeomorphic to $G/H\cong SU(2)/\Z_2$, where $H$ is its isotropy group
\beq
H=\{(\pm h(\lambda),\lambda): \lambda\in U(1)\}.
\eeq
An explicit diffeomorphism $SU(2)/\Z_2\ra\MM$ is given by
$ \{\pm g\}\mapsto (U_*,\omega_*)_{(g,1)}$. 
The action of $G$ on $SU(2)/\Z_2$ induced by this diffeomorphism is 
\beq\label{amoang}
(g,\lambda):\{\pm g'\}\mapsto 
\{\pm h(\lambda)^\dagger g'g\}.
\eeq 

We begin by computing the restriction to $\MM\cong SU(2)/\Z_2$ of the field theoretic Lagrangian
$L=\int_{\R^3}(\LL+\LL_*)$. Since $\LL+\LL_*$ is a quadratic polynomial in time derivatives, this
takes the form
\beq\label{LLL}
L(g,\dot{g})=\frac12\gamma(\dot{g},\dot{g})+A(\dot{g})-M_0
\eeq
where $\gamma$, $A$, $M_0$ are a symmetric $(0,2)$ tensor, a one-form and a function on $SU(2)/\Z_2$
respectively. 
Moreover, $L$ is invariant under the induced action (\ref{amoang}) of $G$ on $SU(2)/\Z_2$, so
$M_0$ is constant and
\beq
\gamma=\Lambda_1(\Sigma_1^2+\Sigma_2^2)+\Lambda_3\Sigma_3^2,\qquad
A=C\Sigma_3
\eeq
where $\Sigma_a$ are the right invariant one forms on $SU(2)$ dual to the basis $\{-\frac{i}{2}\tau_a\}$ for $\su(2)$, and $\Lambda_1,\Lambda_3,C$ are constants. Note that, up to this point, our analysis has used only the axial symmetry of the reference skyrmion $(U_*,\omega_*)$, and so applies equally well
to the exact $B=1$ skyrmion solving the PDE system (\ref{pde}). 

To compute the constants $\Lambda_1,\Lambda_3,C,M_0$, it suffices to evaluate the Lagrangian at $t=0$ on the curves $g_1(t)=\exp(-it\tau_1/2)$ and $g_3(t)=(-iat\tau_3/2)$, where $a\in\R$ is an arbitrary
constant, noting that $L(g_1,\dot{g}_1)=\Lambda_1/2-M_0$ and $L(g_3,\dot{g}_3)=\Lambda_3a^2/2+Ca-M_0$. A lengthy
but straightforward calculation yields
\ben
\Lambda_1&=&\frac{2\pi}{3}\int_0^\infty r^2\sin^2 f\, dr+O(\kappa^2)\\
\Lambda_3&=&\Lambda_1+O(\kappa^2)\\
C&=&\kappa C_*+O(\kappa^2),\qquad C_*=\frac43\int_0^\infty f'\sin f(\frac{\beta}{\pi}\Omega\sin f+\pi r^2\omega_0'\cos f)dr  +O(\kappa^2)\\
M_0&=& 4\pi\int_0^\infty r^2\left(\frac{f'^2}{8}+\frac{\sin^2 f}{4r^2}+\frac{M^2}{4}(1-\cos f)
-\frac{\omega_0'^2}{2}-\frac{\omega_0^2}{2}+\frac{\beta f'\omega_0'\sin^2 f}{2\pi^2r^2}\right)dr+O(\kappa^2).
\een
Of course $M_0$ is the classical rest energy of the static skyrmion. Note that
the symmetry breaking term $\LL_*$ does not break the bi-invariance of the metric $\gamma$ to leading order in $\kappa$; nor does it perturb $M_0$ to leading order. Its only effect is to induce a small
right-invariant one-form on $\MM\cong SU(2)/\Z_2$, tangent to the unbroken isospin symmetry direction. 
For the coupling choice $\beta=34.7$, we find $\Lambda_1=15.242$, $C_*=-5.856$, $M_0=22.505$.

To accommodate the fermionic nature of nucleons, we will quantize motion in $\wt\MM\cong SU(2)$, the double cover of the space of centred skyrmions $\MM$. The wavefunction $\psi$ is a map
$SU(2)\ra\C$ satisfying $\psi(-g)=-\psi(g)$. We recognize in (\ref{LLL}) the Lagrangian describing motion of a particle of unit mass and unit electric charge moving in the Riemannian manifold $(SU(2),\gamma)$ under the influence of a magnetic field $B=\d A$. The Hamiltonian describing the
quantum mechanics of such a system is \cite[p.\ 421]{lanlif}
\beq
H\psi=-\frac12*\d_A*\d_A\psi+M_0\psi
\eeq
where $\d_A=\d-iA$ and $*$ is the Hodge isomorphism on $(SU(2),\gamma)$. For our particular metric $\gamma$ and
connexion $A$,
\beq
H\psi=-\frac1{2\Lambda_1}(\Theta_1^2+\Theta_2^2+\Theta_3^2-2i\kappa C_*\Theta_3)\psi+\left(M_0+\frac{\kappa^2C_*^2}{2\Lambda_1}\right)\psi,
\eeq
where $\{\Theta_a\}$ are the right invariant vector fields on $SU(2)$ dual to $\{\Sigma_a\}$. Since the final term is of order $\kappa^2$, we should discard it. We denote by $\theta_a$ the left invariant vector fields on $SU(2)$ which coincide with $\Theta_a$ at
$\I_2$. To extract the spin and isospin quantum numbers of the eigenstates of $H$, we will re-write it in terms of angular momentum operators. Spatial rotation of the skyrmion
$(U_*,\omega_*)_{(g,1)}$ through angle $\alpha$ about the $x_a$ axis corresponds to {\em right} multiplication of $g$ by $\exp(\alpha i\tau_a/2)$, which is generated by the {\em left} invariant
vector field $-\theta_a$. The corresponding spin operator is $S_a=-i(-i\theta_a)$. Similarly,
isorotation through angle $\alpha$ about the $\pi_3$ axis corresponds to {\em left} multiplication on
$SU(2)$ by $\exp(-\alpha i\tau_3/2)$, which is generated by the {\em right} invariant vector field
$\Theta_3$. The corresponding isospin operator is $J_3=-i\Theta_3$. Now $\Theta_1^2+\Theta_2^2+\Theta_3^2=\theta_1^2+\theta_2^2+\theta_3^2$ and $[\theta_a,\Theta_b]=0$, so
\beq
H=\frac{1}{2\Lambda_1}|\Svec|^2-\frac{\kappa C_*}{\Lambda_1} J_3+M_0,
\eeq
total spin $|\Svec|^2=S_1^2+S_2^2+S_3^2$, $S_3$ and $J_3$ are a compatible set of observables. Hence, by the usual angular momentum algebra, the energy spectrum of the rigidly rotating skyrmion is
\beq
H|s,j_3\rangle=\left(\frac{s(s+1)}{2\Lambda_1}-\frac{\kappa C_*j_3}{\Lambda_1}+M_0\right)|s,j_3\rangle
\eeq
where $s,j_3$ are half integers interpreted as total spin and isospin respectively. The proton corresponds to $|1/2,1/2\rangle$ and the neutron to
$|1/2,-1/2\rangle$. Hence, their masses are
\bea
m_p&=&M_0+\frac{1}{2\Lambda_1}\left(\frac34-\kappa C_*\right), \nonumber \\
m_n&=&M_0+\frac{1}{2\Lambda_1}\left(\frac34+\kappa C_*\right).
\eea

Clearly, for any given parameter value $\beta$, we may choose $\kappa$ so that $m_p$, $m_n$ have the correct
splitting (unless, by some cruel fluke, $C_*= 0$). We must arrange that the mass difference,
as a fraction of the average nucleon mass, equals the experimental value:
\beq
\frac{2(m_n-m_p)}{m_n+m_p}=\frac{2\kappa C_*}{2M_0\Lambda_1+3/4}=0.00137703.
\eeq
For the coupling choice $\beta=34.7$ this requires $\kappa=-0.08075$. A measure of the size of the
perturbation to the skyrmion at this parameter value is given by $\max|\omega_i dx^i|=0.0242$, the maximum length of the spatial part of the $\omega$ meson field over all positions in space (recall this field
vanishes identically for the unperturbed skyrmion). An alternative, more global measure is $\|\omega_idx^i\|/\|\omega_0\|=0.0257$, where $\|\cdot\|$ denotes $L^2$ norm. 

\section{Concluding remarks}

\news

We have proposed a very simple perturbation of the $\omega$ meson Skyrme model which breaks its isospin symmetry to a $U(1)$ subgroup, and is capable of reproducing the neutron-proton mass splitting. 
The perturbed skyrmion's pion and $\omega_0$ fields are unchanged to leading order in the perturbation
parameter $\kappa$, remaining spherically symmetric. The skyrmion acquires an order $\kappa$ azimuthal $\omega_i$ field, however, and so is only axially symmetric. For Sutcliffe's calibration of the
unperturbed model, one must take $\kappa=-0.08075$ to reproduce the correct mass splitting.
One should regard $|\kappa|=0.08075$ as a lower bound on the deformation parameter required, since electromagnetic effects
will partially cancel the desired effect. The perturbation will have implications for pion--$\omega$ scattering processes which we have not addressed. It would be interesting to see whether $|\kappa|=0.08075$ is compatible with the experimental bounds on such processes.

The proposed perturbation is offered in the original
Skyrme spirit: write down something simple that does the job, worry about its microscopic origins later (if at all). In this case, doing the job means inducing a one form on the
moduli space of static skyrmions tangent to the unbroken isospin orbits. This mathematical structure is, presumably, present in some form in any version of the Skyrme model with a neutron-proton mass difference; its derivation in this model is particularly direct and transparent.  

The calculation presented here could be improved. One could numerically solve the Euler-Lagrange equations 
(\ref{pde}), after reduction to axially symmetric fields, instead of working perturbatively. It would be interesting to see whether the pion fields of the skyrmion become
oblate (or prolate). The basic structure of the dynamics on the moduli space would remain unchanged, but the metric would have
only $G$ symmetry (rather than bi-invariance), that is, $\Lambda_1\neq\Lambda_3$,
producing an energy spectrum with a $j_3^2$ term:
\beq
E(s,j_3)=\frac{s(s+1)}{2\Lambda_1}-\frac{Cj_3}{\Lambda_3}+M_0+\frac{C^2}{2\Lambda_3}+\frac{\Lambda_1-\Lambda_3}{2\Lambda_1\Lambda_3}j_3^2.
\eeq
Since chiral symmetry is only softly broken, it would be interesting to study the moduli space dynamics on the orbit of the full spin-isospin group $G'=SU(2)\times SU(2)$, rather than just its unbroken subgroup $G$. The enlarged moduli space $\MM'$ would then be $5$ dimensional, with a $G$ invariant metric and potential (of order $\kappa^2$), presumably minimized on $\MM\subset\MM'$. Calculation of this metric and potential, even perturbatively, seems to be a challenging problem.

\subsection*{Acknowledgements}

The author would like to thank Christoph Adam, Derek Harland and Nick Manton for useful
conversations.


\end{document}